\documentclass[12pt]{iopart}
\usepackage{iopams}
\usepackage{graphicx}
\usepackage[colorlinks=true,dvipdfm]{hyperref}
\usepackage{booktabs}
\usepackage{multirow}
\usepackage[T1]{fontenc}
\hypersetup{
    colorlinks,
    linkcolor=blue,
    citecolor=blue,
    urlcolor=blue,
}
\begin{document}
\title{Bound states of spin-orbit coupled cold atoms in a Dirac
delta-function potential}

\author{Jieli Qin$^1$, Renfei Zheng$^2$, Lu Zhou$^{2,3}$}
\address{$^1$School of Physics and Electronic Engineering, Guangzhou University,
230 Wai Huan Xi Road, Guangzhou Higher Education Mega Center, Guangzhou
510006, China}

\address{$^2$Department of Physics, School of Physics and Electronic Science, East
China Normal University, Shanghai 200241, China}

\address{$^3$Collaborative Innovation Center of Extreme Optics, Shanxi University,
Taiyuan, Shanxi 030006, China}
\ead{\href{mailto:lzhou@phy.ecnu.edu.cn}{lzhou@phy.ecnu.edu.cn}}

\vspace{10pt}
\begin{indented}
\item[] November 2019
\end{indented}

\begin{abstract}
Dirac delta-function potential is widely studied in quantum mechanics
because it usually can be exactly solved and at the same time is useful
in modeling various physical systems.
Here we study a system of delta-potential trapped spin-orbit coupled cold atoms. The
spin-orbit coupled atomic matter wave has two kinds of evanescent modes, one of
which has pure imaginary wavevector and is an ordinary evanescent wave; while
the other with a complex number wave vector is recognized as oscillating
evanescent wave. We identified the eigenenergy spectra and the existence of bound states
in this system. The bound states can be constructed analytically using  the two
kinds of evanescent modes and we found that they exhibit typical features
of stripe phase, separated phase or zero-momentum phase.
In addition to that, the properties of semi-bound states are also discussed, which
is a localized wave packet on a plane wave background.
\end{abstract}

%
\vspace{2pc}
\noindent{\it Keywords}: Spin-orbit coupling, cold atoms, Delta-function potential

\submitto{\JPB}

%
%

\section{Introduction \label{sec:Introduction}}
Spin-orbit (SO) coupling has been widely studied in diverse branches of physics
including nanotechnology \cite{Majumdar2011,Tan2012}, nuclear physics
\cite{Hope1957,Bell1994,Kaiser2007}, optics
\cite{Bliokh2015,Cardano2015,Sala2015}, condensed matter physics
\cite{Bihlmayer2015,Rashba2016} and cold atom physics
\cite{Galitski2013,Zhai2015,Zhang2016}. For a charged particle with non-zero
spin, its spin magnetic momentum will interact with the magnetic field induced
by its movement, thus generating a coupling between its orbital motion and
spin degree of freedom. For neutral cold atom systems, SO-coupling can be
artificially generated via a Raman coupling scheme \cite{Lin2011}. The
occurrence of SO-coupling will greatly enrich the physics of atomic matter
wave \cite{Dalibard2011,Goldman2014,Zhai2015}. Many emergent
phenomena such as spin Hall effect \cite{Sinova2015},
topological insulator \cite{Hasan2010}, Zitterbewegung
\cite{Schrodinger1930,Hestenes2010,LeBlanc2013,Qu2013},
supersolid \cite{Li2017,Wang2017,Liao2018,Zhu2019}, solitons
\cite{Achilleos2013,Xu2013,Zhang2015,Zhu2017}, Beliaev damping \cite{Wu2018}
and spin-dependent atom optics \cite{Jacob2007,Juzeliunas2008PRL,Juzeliunas2008PRA,
Zhou2013,Zhou2015,Zhou2016,Kartashov2016,Kumar2018,Zhang2019,Qin2019,Ji2019,Mossman2019} have been reported.

It was found that in the presence of SO-coupling the atomic system can display
a rich phase diagram in which the ground state wavefunction can favor stripe,
spin separated or zero momentum phase \cite{Lin2011,Wang2010,Ho2011,Li2012,Zheng2013}.
Specifically, the stripe phase can be regarded as a signature of supersolid
\cite{Li2017,Wang2017,Liao2018,Zhu2019} which receives much recent attention. Physical insight
into the properties of eigenfunction of the SO-coupled ultracold atoms can be
gained via bound state solutions. Previous work has solved bound state in a
one-dimensional short-range potential, 
and predicts a type of spin-orbitinduced extra states
\cite{Jursenas2013}. More interestingly,
bound state in the continuum can exist under appropriate trapping potentials
\cite{Kartashov2017}.

In this article we analytically study the bound states of SO-coupled
atomic matterwave in a $\delta$-function potential.
Physical models with $\delta$-function potential play significant roles in
quantum mechanics. Analytical or partially analytical solutions can be deduced
for these models and in the meanwhile they provide physical insight into real
systems. Typical examples include the Kronig-Penney model with $\delta
$-function potential \cite{Kittel2005}, the hydrogen-like atoms and diatomic
molecules \cite{Frost1956}, the inter-atom interaction in ultra cold atomic cloud
\cite{Huang1957,Block2002}, very narrow potential
\cite{Uncu2007,Garrett2011,Lacki2016,Wang2018}, obstacle \cite{Hakim1997,Pavloff2002}
and impurity \cite{Frantzeskakis2002,Seaman2005}. It is well known that a
$\delta$-function potential well supports only one bound state which is
constructed with free particle evanescent waves. For SO-coupled matter wave,
there are three different types of free particle modes: plane wave, ordinary
evanescent wave and oscillating evanescent waves
\cite{Sablikov2007,Zhou2015}. We found that there exist two types of bound
states which can be constructed using the oscillating evanescent and ordinary
evanescent waves, respectively. The bound state constructed with oscillating
evanescent waves has stripe structure on its density profile, while the bound
state with ordinary evanescent waves is a zero momentum wave packet. Due to the
spin-1/2 nature of the system, a $\delta$-function potential well can (but not
always) support
bound states both in ground state and excited state. A separated phase bound
state can then be constructed by superposing the ground and excited bound
state. Besides these bound states, we found that there also exists a kind of
semi-bound state, which is a localized state on a plane wave background. The
interference between the localized state and the plane wave background will
produce a dip (bump) on the density profile for a $\delta$-potential well (barrier).

\section{Model and eigensolution of free SO-coupled cold atomic matter wave \label{sec:ModdelAndFreeParicle}}
\begin{figure}[ptb]
\begin{centering}
\includegraphics{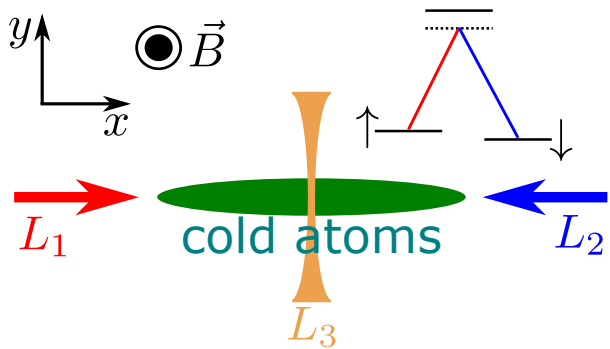}
\par\end{centering}
\caption{
Schematic of the considered system. A quasi-one-dimensional spinor cold atomic cloud
(two Zeeman levels acts as the pseudo-spin states) is  shined by two
$x$-direction counterpropagating laser beams ($L_1$ and $L_2$)
to realize the SO-coupling \cite{Lin2011}. A third far off-resonant tightly focused laser beam
($L_3$) shining from $y$-direction generates a spin-independent delta-function potential \cite{Uncu2007,Garrett2011}.
}
\label{fig:diagram}
\end{figure}

As schematically shown in figure \ref{fig:diagram}, we consider a system of
quasi-one-dimensional SO-coupled cold atoms subjected to a spin-independent
$\delta$-function potential. The SO coupling is realized
by a typical Raman scattering scheme \cite{Lin2011}.
And the spin-independent
$\delta$-potential is generated by a far off-resonant laser beam \cite{Grimm2000}
which is tightly focused at the
center of the atomic cloud \cite{Uncu2007,Garrett2011}.
Such a system can be described by the Hamiltonian%
\begin{equation}
H=H_{0}-V_{0}\delta\left(  x\right)  , \label{eq:Hamiltonian}%
\end{equation}
with $V_{0}$ being the depth of $\delta$-function potential, and $H_{0}$ being
the free particle Hamiltonian of SO-coupled matter wave
\begin{equation}
H_{0}=\left(
\begin{array}{cc}
\frac{\hbar^{2}}{2m}\left(  k_{x}-k_{c}\right)  ^{2} & \hbar\Omega/2\\
\hbar\Omega/2 & \frac{\hbar^{2}}{2m}\left(  k_{x}+k_{c}\right)  ^{2}%
\end{array}
\right)  , \label{eq:H0}%
\end{equation}
which can be implemented with a Raman coupling scheme in cold atom system
\cite{Lin2011}. Here $\hbar$ is the reduced Planck constant, $m$ is the mass
of an atom, $p_{x}=\hbar k_{x}=-i\hbar {\partial}/{\partial x}$ is the
$x$-direction momentum operator, $k_{c}$ signals SO-coupling strength and
$\Omega$ is the Rabi coupling strength. For simplicity, we have assumed that
the interatomic collision interaction is eliminated using the technique of
Feshbach resonance \cite{Chin2010,Timmermans1999}.

Since the total Hamiltonian $H$ equals $H_{0}$ except at the point $x=0$, we
first give a brief discussion on the properties of free particle Hamiltonian
$H_{0}$. The eigenenergy $E$ of $H_{0}$ is given by the equation
\begin{equation}
\left[  E-\frac{\hbar^{2}\left(  k_{x}^{2}+k_{c}^{2}\right)  }{2m}\right]
^{2}-\left(  \frac{\hbar^{2}k_{c}k_{x}}{m}\right)  ^{2}-\left(  \frac
{\hbar\Omega}{2}\right)  ^{2}=0. \label{eq:eigenvalues}%
\end{equation}
For a given energy $E$, this quartic algebraic equation has four solutions
\begin{equation}
k_{x}=\pm\sqrt{k_{c}^{2}+\frac{2mE}{\hbar^{2}}\pm\frac{\sqrt{8mEk_{c}%
^{2}+m^{2}\Omega^{2}}}{\hbar}}. \label{eq:kx}%
\end{equation}
These four solutions can generally be written as $k_{x}%
=\beta + i\alpha$, with $\alpha$ and $\beta$ being  two real numbers.
In the wavefunction $\textrm{e}^{i k_x x}=\textrm{e}^{-\alpha x} \textrm{e}^{i\beta x}$,
the real part $\beta$ of wavevector $k_x$ contributes a plane wave factor,
while the imaginary part $\alpha$ contributes an exponential decay factor. Thus,
depending on the values of $\alpha$ and $\beta$, the
corresponding free particle waves can be divided into three
types: (1) Plane wave with $\alpha=0$, $\beta\neq0$; (2) Ordinary evanescent
wave with $\alpha\neq0$, $\beta=0$; and (3) Oscillating evanescent wave when
$\alpha,\beta\neq0$.

From equation (\ref{eq:eigenvalues}), eigenenergy can be further split into two
spectrum branches (referred as upper and lower branch)
\begin{equation}
E_{\pm}\left(  k_{x}\right)  =\frac{\hbar^{2}\left(  k_{x}^{2}+k_{c}%
^{2}\right)  }{2m}\pm\sqrt{\left(  \frac{\hbar^{2}k_{c}k_{x}}{m}\right)
^{2}+\left(  \frac{\hbar\Omega}{2}\right)  ^{2}}, \label{eq:dispersion}%
\end{equation}
with the corresponding eigenstates%
\begin{equation}
\Psi_{\pm}=\chi_{\pm}\left(  k_{x}\right)  e^{ik_{x}x}=C_{\pm}\left(\!\!
\begin{array}{c}
\zeta_{\pm}\\
1
\end{array}
\!\!\right)  e^{ik_{x}x}, \label{eq:eigenvector_p}%
\end{equation}%
in which $\zeta_{\pm}=-\left(  2\hbar k_{c}k_{x}\mp\sqrt{4\hbar^{2}k_{c}%
^{2}k_{x}^{2}+m^{2}\Omega^{2}}\right)  /m\Omega$ and
$C_{\pm}=1/\sqrt{1+\left\vert \zeta_{\pm}\right\vert ^{2}}$ is the normalization constant.

\begin{figure}[ptb]
\begin{centering}
\includegraphics{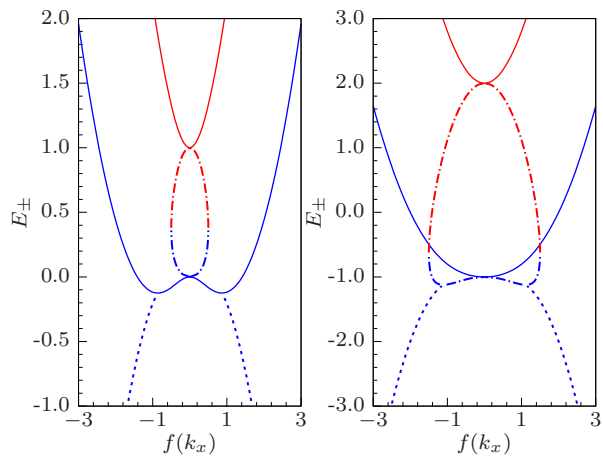}
\par\end{centering}
\caption{Free particle energy spectra of SO-coupled atomic matter
waves. Left panel: strong SO-coupling case with parameters $k_{c}^{2}%
>m^{2}\Omega^{2}/\left(  4\hbar\right)  $ ($k_{c}=1,\Omega=1$). Right panel:
weak SO-coupling case with parameters $k_{c}^{2}<m^{2}\Omega^{2}/\left(
4\hbar\right)  $ ($k_{c}=1,\Omega=3$). The upper branch spectrum $E_{+}$ is
plotted in red color, while the lower branch $E_{-}$ is plotted in blue color.
The plane, ordinary evanescent and oscillating evanescent wave modes are
plotted with solid, dash-dot and dash lines respectively. For plane wave mode,
$k_{x}$ is a real number, $x$-axis is simply set to $f\left(  k_{x}\right)
=k_{x}$; for ordinary evanescent wave mode, $k_{x}$ has no real part,
$x$-axis is set to its imaginary part $f\left(  k_{x}\right)  =\textrm{Im}%
\left(  k_{x}\right)  $; and for oscillating evanescent wave mode, $k_{x}%
=\pm\beta\pm i\alpha$ is a complex number, $x$-axis is set to $f\left(
k_{x}\right)  =\textrm{sgn}\left[  \textrm{Im}\left(  k_{x}\right)  \right]
\cdot\left\vert k_{x}\right\vert $ (note that in such a way the points for
$+\beta$ and $-\beta$ will overlap with each other, thus the seemingly two
curves in the figure are in fact four curves). Other parameters used are
$m=\hbar=1$. }%
\label{fig:EnergySpectrum}%
\end{figure}

\begin{table*}[ptb]
\begin{centering}
\begin{tabular}{llccccccccccccc}
\toprule
\multirow{3}{*}{Energy range} &  & \multicolumn{6}{c}{Strong coupling } &  & \multicolumn{6}{c}{Weak coupling }\tabularnewline
\cmidrule{3-8} \cmidrule{10-15}
&  & \multicolumn{2}{c}{P} & \multicolumn{2}{c}{Ev} & \multicolumn{2}{c}{OE } &  & \multicolumn{2}{c}{P} & \multicolumn{2}{c}{Ev} & \multicolumn{2}{c}{OE}\tabularnewline
\cmidrule{3-8} \cmidrule{10-15}
&  & U & L & U & L & U & L &  & U & L & U & L & U & L\tabularnewline
\cmidrule{1-1} \cmidrule{3-8} \cmidrule{10-15}
$\left[E_{+}\left(0\right),+\infty\right]$ &  & 2 & 2 &  &  &  &  &  & 2 & 2 &  &  &  & \tabularnewline
\cmidrule{1-1} \cmidrule{3-8} \cmidrule{10-15}
$\left[E_{-}\left(i\alpha_{0}\right),E_{+}\left(0\right)\right]$ &  &  & 2 & 2 &  &  &  &  &  & 2 & 2 &  &  & \tabularnewline
\cmidrule{1-1} \cmidrule{3-8} \cmidrule{10-15}
$\left[E_{-}\left(0\right),E_{-}\left(i\alpha_{0}\right)\right]$ &  &  & 2 &  & 2 &  &  &  &  & 2 &  & 2 &  & \tabularnewline
\cmidrule{1-1} \cmidrule{3-8} \cmidrule{10-15}
$\left[E_{-}\left(k_{0}\right),E_{-}\left(0\right)\right]$ &  & 4 &  &  &  &  &  &  &  &  &  & 4 &  & \tabularnewline
\cmidrule{1-1} \cmidrule{3-8} \cmidrule{10-15}
$\left[-\infty,E_{-}\left(k_{0}\right)\right]$ &  &  &  &  &  &  & 4 &  &  &  &  &  &  & 4\tabularnewline
\bottomrule
\end{tabular}
\par\end{centering}
\caption{Plane (P), ordinary evanescent (Ev), and oscillating evanescent (OE)
mode numbers of SO-coupled atomic matter wave in
different energy ranges. Strong coupling means $k_{c}^{2}>m^{2}\Omega
^{2}/\left(  4\hbar\right)  $, while weak coupling means $k_{c}^{2}%
<m^{2}\Omega^{2}/\left(  4\hbar\right)  $. Letters \textquotedblleft
U\textquotedblright\ and \textquotedblleft L\textquotedblright\ are used to
label the upper and lower branch of the spectrum. In the table $\alpha_{0}$
and $k_{0}$ are $\alpha_{0}=m\Omega/\left(  2\hbar k_{c}\right)  $,
$k_{0}=\sqrt{k_{c}^{2}-m^{2}\Omega^{2}/\left(  4\hbar^{2}k_{c}^{2}\right)  }$.
The expression of function $E_{\pm}$ is given by formulae (\ref{eq:dispersion}%
) in the text. }%
\label{tab:ModeNumbers}%
\end{table*}

Equation (\ref{eq:kx}) indicates that there are four states for any energy $E$, the
eigenenergy dispersion display two typically different structures depending on
the SO-coupling strength:

(i) The strong SO-coupling case with $k_{c}^{2}>m^{2}\Omega^{2}/4\hbar$. In
this case, when $E>E_{+}\left(  0\right)  =\hbar^{2}k_{c}^{2}/2m+\hbar
\Omega/2$, both the upper and lower branches support two plane wave states
with $k_{x}$ real. In the region $E\in\left[  E_{-}\left(  0\right)
=\hbar^{2}k_{c}^{2}/2m-\hbar\Omega/2,E_{+}\left(  0\right)  \right]  $, there
are two plane wave states in the lower spectrum branch while the other two
are ordinary evanescent states coming from either the upper branch when $E>E_{-}\left(
i\alpha_{0}\right)  =\left(  4\hbar^{2}k_{c}^{4}-m^{2}\Omega^{2}\right)
/8mk_{c}^{2}$ $\left(  \alpha_{0}=m\Omega/2\hbar k_{c}\right)  $ or lower
branch when $E<E_{-}\left(  i\alpha_{0}\right)  $. For $E\in\left[
E_{-}\left(  k_{0}\right)  =-m\Omega^{2}/8k_{c}^{2},E_{-}\left(  0\right)
\right]  $ with $k_{0}=\sqrt{k_{c}^{2}-m^{2}\Omega^{2}/\left(  4\hbar^{2}%
k_{c}^{2}\right)  }$ there are four plane wave states in the lower branch.
Four oscillating evanescent states possess minimum energies with
$E<E_{-}\left(  k_{0}\right)  $ and they are linked to the energy minimum of
the upper energy spectra at the points $\left\vert k_{x}\right\vert =k_{0}$.

(ii) The weak SO-coupling case with $k_{c}^{2}<m^{2}\Omega^{2}/4\hbar$. It
differs from the strong coupling case only in the energy region $\left[
E_{-}\left(  k_{0}\right)  ,E_{-}\left(  0\right)  \right]  $, in which all
four eigenstates come from the lower branch are ordinary evanescent waves
with $k_{x}$ imaginary.

In order to better understand the properties of these eigenstates, we plot in
figure \ref{fig:EnergySpectrum} the energy spectra as a function of $k_{x}$ for
$k_{c}=1$ and two typical values of $\Omega$, signaling the strong coupling
and weak coupling case respectively. The corresponding numbers of plane,
evanescent and oscillating evanescent modes in different energy ranges are
also summarized in Table \ref{tab:ModeNumbers}.

\section{Bound states and semi-bound states with delta-function potential \label{sec:BoundStates}}
The bound state of a $\delta$-function potential well $V(x)=-V_0 \delta(x)$
can be constructed using the free particle modes by matching the boundary
conditions at $x=0$. Because the plane wave mode extends to infinity, it can
not be used to construct a bound state. While ordinary and oscillating
evanescent wave modes decay to zero when $x$ approaches $+\infty$ or
$-\infty$, they are the candidates for bound state constructing. So based on
the discussion in section \ref{sec:ModdelAndFreeParicle}, it is concluded that
bound states can exist in energy range $\left[-\infty,E_{-}\left(k_0\right)\right]$
for both strong and weak SO-coupling, and in energy range $\left[E_{-}\left(k_0\right),
E_{-}\left(0\right) \right]$ for only weak SO-coupling, since in these energy ranges
there exist the candidate modes. And in energy range $\left[E_{-}\left(0\right),
E_{+}\left(0\right)\right]$, the plane wave and ordinary evanescent wave modes exist
simultaneously, this gives a chance to construct a kind of semi-bound state, which
shows as a localized wave packet on plane wave background.
These states will be discussed in the rest content of this section one by one.

In the energy range $E<E_{-}\left(  k_{0}\right)  $, as
having been discussed above there exist four oscillating evanescent modes and
bound states can be constructed using them. One can find that the wave vector
(\ref{eq:kx}) of these four modes have symmetrical form $k_{x}=\pm\beta\pm
i\alpha$ and we label them as $k_{1}=\beta+i\alpha$, $k_{2}=-\beta+i\alpha$,
$k_{3}=\beta-i\alpha$, $k_{4}=-\beta-i\alpha$, so the bound state can be
written as
\begin{equation}
\Psi_{b}\!=
\!\!\left\{\!\!\!\!
\begin{array}{l@{\,}l}
\left[ A_{1}\chi_{-}\!\left(  k_{1}\right)  \!e^{i\beta x}\!+\!A_{2}\chi
_{-}\!\left(  k_{2}\right)  \!e^{-i\beta x}  \right]  \!e^{-\alpha x}, & x>0,\\
\left[ A_{3}\chi_{-}\!\left(  k_{3}\right)  \!e^{i\beta x}\!+\!A_{4}\chi
_{-}\!\left(  k_{4}\right)  \!e^{-i\beta x} \right]  \!e^{\alpha x}, & x<0,
\end{array}
\right.\!\!
\label{eq:bound_OE}%
\end{equation}
with $A_{1},A_{2},A_{3},A_{4}$ and eigenenergy $E$ {(}note that $k_{1,2,3,4}$
are determined by $E$ according to formula
(\ref{eq:kx}){)} to be determined by normalization constraint
\begin{equation}
\int_{-\infty}^{\infty}\left\vert \Psi_{b}\left(  x\right)  \right\vert
^{2}dx=1,\label{eq:normalization}%
\end{equation}
and boundary conditions at $x=0$: continuity of the wave function
\begin{equation}
\Psi_{b}|_{0+}=\Psi_{b}|_{0-},\label{eq:boundary_1}%
\end{equation}
with
\begin{eqnarray}
\Psi_{b}|_{0+}  & =A_{1}\chi_{-}\left(  k_{1}\right)  +A_{2}\chi_{-}\left(
k_{2}\right)  ,\\
\Psi_{b}|_{0-}  & =A_{3}\chi_{-}\left(  k_{3}\right)  +A_{4}\chi_{-}\left(
k_{4}\right)  ,
\label{eq_wavefunction}
\end{eqnarray}
and jump of the first-order derivation caused by divergence of $\delta
$-function potential
\begin{equation}
\left.\frac{\partial\Psi_{b}}{\partial x}\right|_{0+}-\left.\frac{\partial\Psi_{b}}{\partial
x}\right|_{0-}=-\frac{2mV_{0}}{\hbar^{2}}\Psi_{b}\left(  x=0\right)
,\label{eq:boundary_2}%
\end{equation}
with
\begin{eqnarray}
\left.\frac{\partial\Psi_{b}}{\partial x}\right|_{0+}  & =ik_{1}A_{1}\chi_{-}\left(
k_{1}\right)  +ik_{2}A_{2}\chi_{-}\left(  k_{2}\right)  ,\\
\left.\frac{\partial\Psi_{b}}{\partial x}\right|_{0-}  & =ik_{3}A_{3}\chi_{-}\left(
k_{3}\right)  +ik_{4}A_{4}\chi_{-}\left(  k_{4}\right)  .
\label{eq_derivative}%
\end{eqnarray}

\begin{figure*}[ptb]
\centering{}\includegraphics{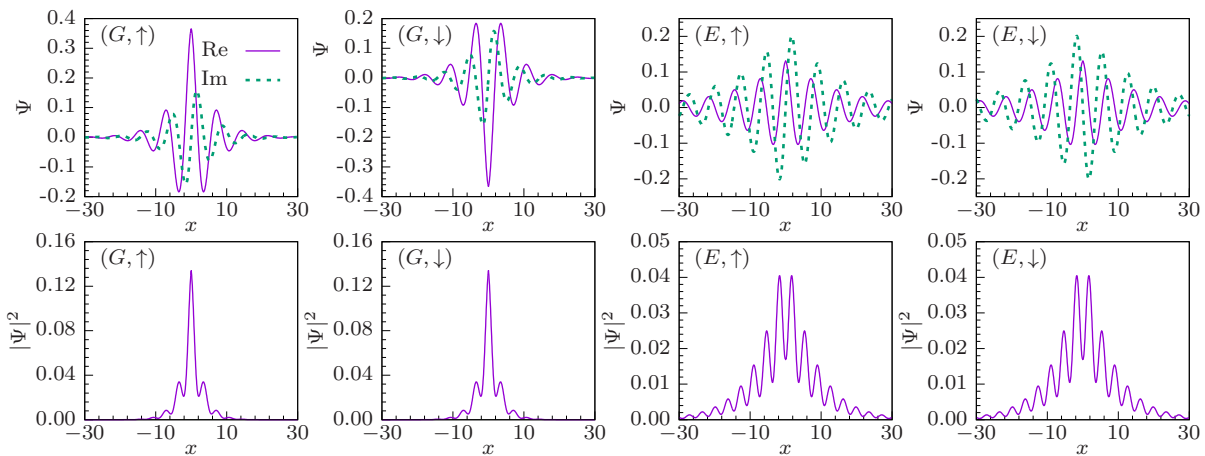}\caption{Oscillating evanescent wave
bound states. Ground (\textquotedblleft G\textquotedblright) and Excited
(\textquotedblleft E\textquotedblright) states of SO-coupled matter wave in
$\delta$-function potential well are plotted for parameters $V_{0}%
=0.1,k_{c}=1,\Omega=1.$ Top panels are real (solid) and imaginary (dashed)
parts of the wave function $\Psi$ with different spins (labeled with
\textquotedblleft$\uparrow$\textquotedblright\ and \textquotedblleft%
$\downarrow$\textquotedblright). Bottom panels are the corresponding densities
$\left\vert \Psi\right\vert ^{2}$. The energies of the ground and excited states
are $-0.1391$ and $-0.1267$ respectively. Natural unit $m=\hbar=1$ is applied.
}%
\label{fig:StripeStates}%
\end{figure*}

Solving equations (\ref{eq:boundary_1}) and (\ref{eq:boundary_2}), one can have two
solutions (the lower energy one is the ground state and the other one is an
excited state) fulfill properties $\left\vert A_{1}\right\vert =\left\vert
A_{2}\right\vert =\left\vert A_{3}\right\vert =\left\vert A_{4}\right\vert $,
indicating that the bound state is a spin symmetric state with $\left\vert
\Psi_{b,\uparrow}\right\vert ^{2}=\left\vert \Psi_{b,\downarrow}\right\vert
^{2}$. From equation (\ref{eq:bound_OE}) one can understand that the interference
between $e^{i\beta x}$ and $e^{-i\beta x}$ terms will produce an interference
stripe on the density profile of bound states. This type of bound state has
very similar properties as the stripe phase state in free space or
harmonically trapped SO-coupled matter wave \cite{Li2012}. In figure
\ref{fig:StripeStates}, examples of such bound states are plotted for
parameters $V_{0}=0.1,k_{c}=1,\Omega=1$, clearly demonstrating spin mixed
stripe phase structure. We also note that no-node theorem does not hold here
because of SO-coupling \cite{Wu2009,Zhou2013JPB}.

\begin{figure}[ptb]
\begin{centering}
\includegraphics{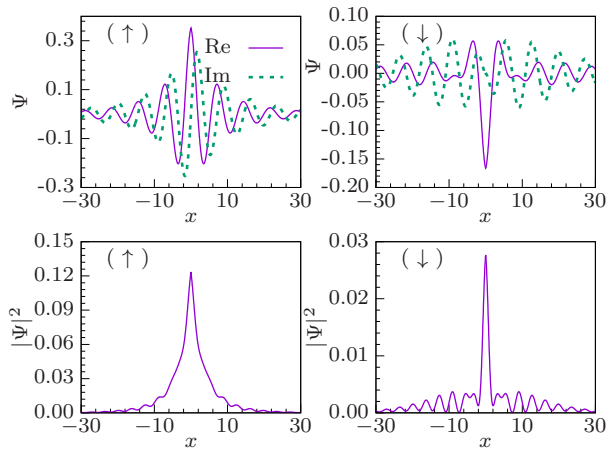}
\par\end{centering}
\caption{Separated phase bound state as a superposition of ground and excited
states, $\left(  \Psi_{b,G}+\Psi_{b,E}\right)  /\sqrt{2}$. Top panels are the
real (solid) and imaginary (dashed) parts of the spin-$\uparrow$ and
spin-$\downarrow$ wave functions. Bottom panels are the corresponding
densities $\left|  \Psi\right|  ^{2}$. Parameters are the same as in figure
\ref{fig:StripeStates}. }%
\label{fig:SpinSeparatedState}%
\end{figure}

The bound states can also be constructed via the linear superposition of the
ground and excited states. In figure \ref{fig:SpinSeparatedState}, the
superposition state $\Psi_{b,G+E}=\left(  \Psi_{b,G}+\Psi_{b,E}\right)
/\sqrt{2}$ is plotted which is a spin-$\uparrow$ component dominated state.
Similarly a spin-$\downarrow$ component dominated state can also be
constructed by superposing $\Psi_{b,G}$ and $\Psi_{b,E}$ with opposite phase
$\Psi_{b,G-E}=\left(  \Psi_{b,G}-\Psi_{b,E}\right)  /\sqrt{2}$. This resembles
the separated phase discussed in \cite{Li2012} with nonzero spin polarization.
But, it should be noted that because the ground and excited states have different
eigenenergies, these superposition states are not stationary states of the system.

\begin{figure}[ptb]
\begin{centering}
\includegraphics{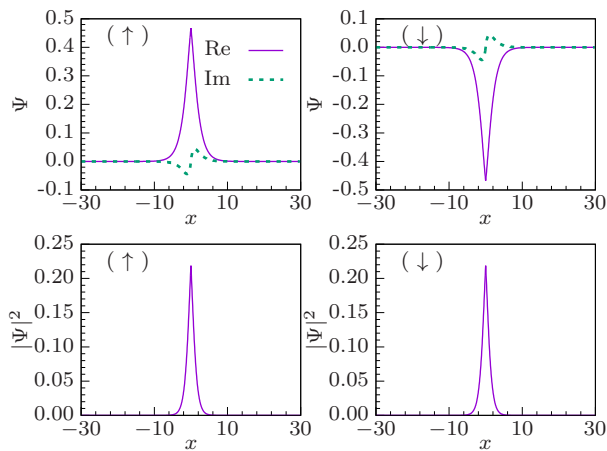}
\par\end{centering}
\caption{Ordinary evanescent wave bound state. Ground state of SO-coupled
matter wave in $\delta$-function potential well is plotted for parameters
$V_{0}=0.25,k_{c}=1,\Omega=3$. Top panels are the real (solid) and imaginary
(dashed) parts of the spin-$\uparrow$ and spin-$\downarrow$ wave functions.
Bottom panels are the corresponding densities $\left\vert \Psi\right\vert
^{2}$. The energy of this state is $-1.0625$. Natural unit $m=\hbar=1$ is
applied. }%
\label{fig:ZeroMomentumState}%
\end{figure}

The bound states can also exist in the energy region $\left[  E_{-}\left(
k_{0}\right)  ,E_{-}\left(  0\right)  \right] $ for a weak SO-coupling
($k_c^2<m^2\Omega^2/(4\hbar)$). In such a case, there exist four ordinary evanescent modes
for a given energy $E$, the corresponding wavevectors (\ref{eq:kx})
are pure imaginary and have form $k_{1,3}=\pm i\kappa_{1},k_{2,4}=\pm i\kappa_{2}$.
The bound state can then be written in the following form
\begin{equation}
\Psi_{b}\!=\!\!\left\{\!\!\!\!
\begin{array}{l@{\,}l}
A_{1}\chi_{-}\left(  k_{1}\right)  e^{-\kappa_{1}x}\!+\!A_{2}\chi_{-}\left(
k_{2}\right)  e^{-\kappa_{2}x}, & x>0,\\
A_{3}\chi_{-}\left(  k_{3}\right)  e^{\kappa_{1}x}\!+\!A_{4}\chi_{-}\left(
k_{4}\right)  e^{\kappa_{2}x}, & x<0,
\end{array}
\right.
\label{eq:bound_Ev}%
\end{equation}
which decays exponentially and is very similar to the bound states in the
SO-uncoupled case. Applying the boundary conditions and normalization
constraints {(}similar to equaitons (\ref{eq:normalization},
\ref{eq:boundary_1}, \ref{eq:boundary_2}){)} one can solve the bound states.
An example is given in figure \ref{fig:ZeroMomentumState}, the bound state is
spin symmetric and can be viewed as the ``zero momentum'' states discussed in \cite{Li2012}.

\begin{figure}[ptb]
\begin{centering}
\includegraphics{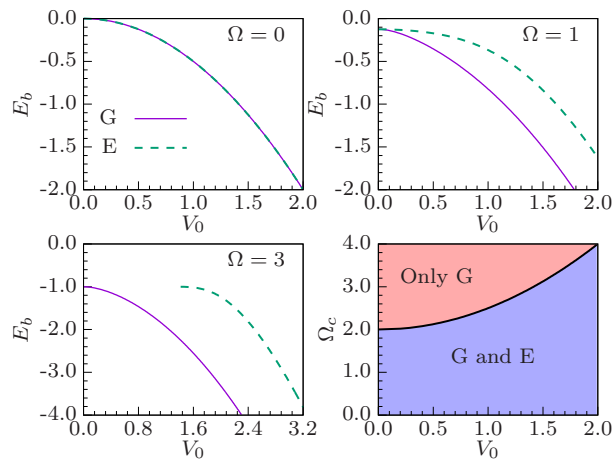}
\par\end{centering}
\caption{
$\delta$-function potential well bounded spectrum. The binding energies
of the ground(``G'') and excited(``E'') states are plotted as a function of potential
well depth $V_0$ for parameters $\Omega=0,1,3$ in the top two and bottom left panels
respectively. In the bottom right panel, the critical value of Rabi coupling
$\Omega_c$ for the disappearing of the excited state is plotted as a function of potential
well depth $V_0$, the solid black line. Below this line, in the light blue color filled
area, there exist both the ground and excited states. While, above this line, in the
light red color filled area, only the ground state can exist. In all the panels, SO-coupling
strength is set to $k_c=1$ and natural unit $m=\hbar=1$ is applied.
}
\label{fig:BoundSpectrum}%
\end{figure}

We also examined the spectrum with $\delta$-function potential well. In the
top two and bottom left panels of figure \ref{fig:BoundSpectrum}, the binding
energies of the ground and excited bound states are plotted as a function of $\delta$-function
potential well depth $V_0$ for Rabi coupling strength $\Omega=0,1$ and $3$. When $\Omega=0$,
the two spin components are not coupled with each other, each component can be separately treated as a
usual $\delta$-function potential problem (except that the momentum is shifted by $\pm \hbar k_c$),
thus there exist two bound states with degenerate energy $E_{b;G}=E_{b;E}=-mV_0^2/(2\hbar^2)$.
When $\Omega \neq 0$, this degeneracy
is eliminated. For a small value of $\Omega=1$ (or in other words, a strong SO-coupling since
$k_c^2 > m^2\Omega^2/4\hbar$ is fulfilled), both the ground and excited states can exist regardless
of the value of $V_0$. Even when $V_0 \rightarrow 0$ (approaching the free particle limit), the
system has two solutions corresponding to the two minimums of the lower dispersion branch. However,
for a large value of $\Omega=3$ (weak SO-coupling since $k_c^2 < m^2\Omega^2/4\hbar$), the excited state
can be lifted so high that a shallow potential well can no longer trap it. Thus, we see that for $V_0$
smaller than a critical value, the excited state disappears. This also coincides with the fact that
for weak SO-coupling the lower dispersion branch of the free particle spectrum only has one minimum. And in the
bottom right panel, we show the critical value of Rabi coupling strength $\Omega_c$ for the disappearing
of the excited state as a function of $V_0$, see the solid black line. Below this critical line, both the ground and
excited states can exist. While above this line, the potential well can no longer support an excited bound
state, only the ground state can exist. In this panel, we also noticed that when $V_0 \rightarrow 0$, the
critical Rabi coupling $\Omega_c \rightarrow 2$ which is just the demarcation point between strong and weak
SO-coupling strength ($\sqrt{4\hbar k_c^2/m^2}=2$). This also agrees with the above free particle
limit discussion.

\begin{figure}[ptb]
\begin{centering}
\includegraphics{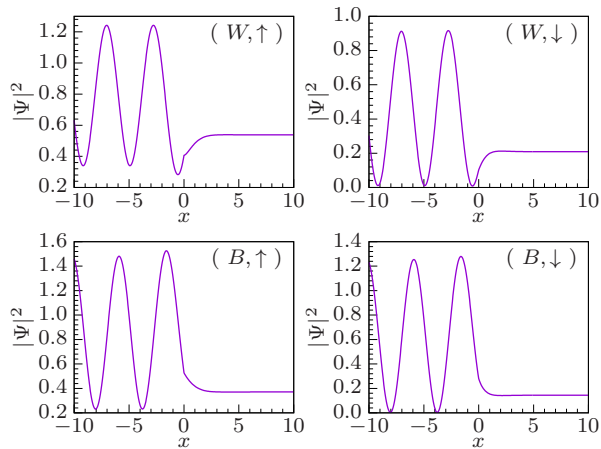}
\par\end{centering}
\caption{Semi-bound states of SO-coupled matter wave in $\delta
$-function potential well (``W'', top panels) and barrier (``B'', bottom
panels). Density profiles of spin-$\uparrow$ and spin-$\downarrow$ components
are plotted. Parameters used are $E_{0}=-0.9$, $V_{0}=\pm0.25$ ($+0.25$ for a
potential well, while $-0.25$ for a potential barrier), $k_{c}=1$, $\Omega=3$
and $m=\hbar=1$. }%
\label{fig:SemiBoundState}%
\end{figure}

In the energy region $\left[  E_{-}\left(  0\right)  ,E_{+}\left(  0\right)
\right]  $, for a given energy $E$ there are two ordinary evanescent modes and two
plane wave modes, the corresponding wavevectors (\ref{eq:kx})
have form $k_{1,3}=\pm i \kappa$ and $k_{2,4}=\pm k$. One can then construct a
semi-bound state as follows
\begin{equation}
\Psi_{sb}=\Psi_{P}+\Psi_{E},\label{eq:semiBound}%
\end{equation}
where $\Psi_{P}$ is a plane wave background consisting of incident,
transmission and reflection waves (their amplitudes are $1,t,r$ respectively)
\begin{equation}
\Psi_{P}\!=\!\!\left\{\!\!\!\!
\begin{array}{l@{\,}l}
t\chi_{-}\left(  k_{2}\right)  e^{ikx}, & x>0,\\
\chi_{-}\left(  k_{2}\right)  e^{ikx}+r\chi_{-}\left(  k_{4}\right)
e^{-ikx}, & x<0,
\end{array}
\right.
\label{eq:PlaneBackground}%
\end{equation}
and $\Psi_{E}$ is a localized wave packet constructed using the two
ordinary evanescent modes
\begin{equation}
\Psi_{E}\!=\!\!\left\{\!\!\!\!
\begin{array}{l@{\,}l}
A_{1}\chi_{\pm}\left(  k_{1}\right)  e^{-\kappa x}, & x>0,\\
A_{3}\chi_{\pm}\left(  k_{3}\right)  e^{\kappa x}, & x<0.
\end{array}
\right.
\label{eq:Localized}%
\end{equation}
Here $t,r$ and $A_{1},A_{3}$ are parameters to be determined by boundary
conditions. We note that such a semi-bound state is actually a scattering
state, thus it can not only exist for a $\delta$-potential well but also a
$\delta$-potential barrier. In figure \ref{fig:SemiBoundState}, examples of such
semi-bound states are shown for parameters $k_{c}=1,$$\Omega=3$ and $V_{0}%
=\pm0.25$ ($+0.25$ for a potential well, while $-0.25$ for a potential
barrier). The density profiles of both spin components are plotted in this
figure. In the left-half space ($x<0$), the interference between incident and
reflected waves produces an interference fringe. In the right-half space
($x>0$), the transmission wave produces a flat density background. And around
the location of $\delta$-potential ($x=0$), a dip of density on plane wave
ground can be observed for a potential well, while a bump is observed for a
potential barrier. This semi-bound state represents the coupling between the
bound state and the plane wave propagating state, thus bound states in the
continuum don't exist in the present single-particle system \cite{Hsu2016}.

\section{Summary}
In summary, we studied the bound and semi-bound states of SO-coupled
matter wave in a $\delta$-function potential. We found that there are two
kinds of bound state in the system, one of which is a stripe one constructed
using oscillating evanescent wave, while the other one constructed using
ordinary evanescent wave is an ordinary one having the similar feature as the
SO-uncoupled case. For SO-coupled matter wave, a $\delta$-potential well can
(but not always) support both a ground and an excited bound state. By
superposing these two states, a separated phase state can also be constructed.
Besides the bound states, there also exists a kind of semi-bound state (a
localized wave packet on a plane wave background). For a $\delta$-function
potential well, a dip emerges on the plane wave background. While for a
$\delta$-function potential barrier, a bump is formed on the plane wave background.

\ack
This work is supported by the National Natural Science Foundation of China
(Grant Nos. 11904063, 11847059, 11374003, and 11574086).

\end{document}